\documentclass[11pt]{article}

\usepackage[utf8]{inputenc}
\usepackage[T1]{fontenc}
\usepackage{times}
\usepackage{latexsym}
\usepackage{microtype}
\usepackage{inconsolata}
\usepackage{graphicx}
\usepackage{amsmath}
\usepackage{amsfonts}
\usepackage{amssymb}

\usepackage[numbers]{natbib}

\usepackage[colorlinks=true,linkcolor=blue,citecolor=blue,urlcolor=blue]{hyperref}

\usepackage[margin=1in]{geometry}

\title{Beyond Passive Viewing: A Pilot Study of a Hybrid Learning Platform Augmenting Video Lectures with Conversational AI}

\author{
Mohammed Abraar \quad Raj Abhijit Dandekar \quad Rajat Dandekar \quad Sreedath Panat \\
Vizuara AI Labs \\
\texttt{\{abraar,raj,rajatdandekar,sreedath\}@vizuara.com}
}

\begin{document}
\maketitle

\begin{abstract}
The exponential growth of AI education has brought millions of learners to online platforms, yet this massive scale has simultaneously exposed critical pedagogical shortcomings. Traditional video-based instruction, while cost-effective and scalable, demonstrates systematic failures in both sustaining learner engagement and facilitating the deep conceptual mastery essential for AI literacy. We present a pilot study evaluating a novel hybrid learning platform that integrates real-time conversational AI tutors with traditional video lectures. Our controlled experiment (N = 58, mean age M = 21.4, SD = 2.8) compared traditional video-based instruction with our AI-augmented video platform. This study employed a sequential within-subjects design where all participants first completed the traditional video condition followed by the AI-augmented condition, providing direct comparisons of learning outcomes. We measured learning effectiveness through immediate post-tests and delayed retention assessments (2-week delay). Results suggest improvements in learning performance: immediate post-test performance showed a large effect size (d = 1.505) with participants scoring 8.3 points higher after AI-augmented instruction (91.8 vs 83.5 out of 100, $p < .001$). Behavioral analytics revealed increased engagement duration (71.1\% improvement with AI tutoring) in the experimental group. This pilot study provides preliminary evidence that conversational AI tutors may enhance traditional educational delivery, suggesting a potential avenue for developing scalable, adaptive learning systems.
\end{abstract}

\section{Introduction}

The global surge in Artificial Intelligence (AI) education has redefined the landscape of learning. Online platforms, MOOCs, and digital academies have collectively attracted millions of learners worldwide, democratizing access to AI knowledge \citep{ref11}. However, this massive scale of participation has simultaneously revealed fundamental pedagogical shortcomings. Traditional video-based instruction, which remains the backbone of online education, is cost-effective and scalable, yet it frequently produces passive learning experiences that fail to sustain engagement or facilitate durable conceptual mastery \citep{ref3,ref6,ref10}. Decades of research in the learning sciences have consistently demonstrated that passive exposure to content results in limited knowledge retention, poor conceptual transfer, and a higher likelihood of learner disengagement \citep{ref7,ref8,ref13}.

To overcome these challenges, education researchers have explored interactive learning modalities. Theories of active learning and constructivist pedagogy emphasize that learners build deeper understanding when they are actively engaged in dialogue, problem-solving, and application-driven exploration \citep{ref3,ref10}. In practice, however, such approaches are difficult to scale in large online environments, where the ratio of learners to human instructors is disproportionately high \citep{ref1,ref9}. This has led to growing interest in the application of artificial intelligence technologies to provide scalable interactivity in digital education.

Over the past decade, AI-driven tutoring systems, conversational agents, and adaptive learning platforms have shown promising results in improving learning outcomes. Studies demonstrate that AI tutors can approximate one-to-one tutoring effects by offering immediate feedback, scaffolding, and personalized pacing \citep{ref19}. Conversational AI in particular has been found to support deeper questioning, clarify misconceptions, and increase motivation in online courses \citep{ref8,ref13,ref9}. Experimental evidence also suggests that AI-augmented instruction can significantly improve knowledge retention and conceptual transfer compared to video-only formats \citep{ref18,ref20}. Furthermore, the integration of natural language dialogue into learning systems has opened possibilities for hybrid approaches that combine the scalability of videos with the adaptivity of AI \citep{ref5}. Despite these advances, most existing implementations remain limited in scale, constrained to controlled classroom studies or specialized adaptive platforms \citep{ref1,ref12,ref17}.

The question of whether conversational AI tutors can transform massive, video-centered education ecosystems has yet to be systematically investigated \citep{ref11,ref12,ref17}. Videos remain central to how AI education is delivered globally, but the gap between passive consumption and active learning persists, with large-scale MOOC analyses showing short attention spans and sharp drop-offs in engagement during videos \citep{ref21,ref22,ref23}. This creates an urgent need to explore how conversational AI can be embedded within video-based instruction to provide scalable, interactive, and pedagogically effective experiences aligning with evidence that active and dialogic engagement outperforms passive viewing for durable learning \citep{ref24,ref25,ref26,ref27,ref28}.

In this paper, we present a pilot study of a novel hybrid learning paradigm that integrates conversational AI tutors directly into traditional video lectures. Rather than replacing videos, our system augments them with real-time interactive dialogue, enabling learners to ask clarifying questions, receive adaptive feedback, and test conceptual understanding as they progress through video content. Grounded in the principle that interactive/constructive engagement fosters deeper processing than passive or merely active behaviors \citep{ref26}, this design leverages recent advances in natural-language tutoring and intelligent tutoring systems to provide adaptive support at scale \citep{ref5,ref19,ref28}. Early studies that embed chatbots or pedagogical agents alongside instructional media show promise for improved motivation, engagement, and metacognitive gains, while surfacing important challenges in privacy, over-reliance, and evaluation rigor \citep{ref8,ref9,ref10,ref11,ref12,ref20}. Our approach operationalizes these insights by coupling in-video, just-in-time questioning, adaptive hints, and formative checks with conversational scaffolding aligned to course objectives \citep{ref1,ref3,ref5,ref17,ref19}.

We conducted a controlled pilot experiment (N = 58) to evaluate the effectiveness of this AI-augmented video platform. Learners engaged with both traditional video instruction and AI-augmented videos in a sequential within-subjects design, enabling direct comparison of the impact of conversational interactivity. We measured learning outcomes across immediate performance and delayed retention, while also analyzing behavioral engagement patterns. Our design choices are informed by robust findings that (i) active learning reliably improves achievement and reduces failure rates compared with lecture \citep{ref24,ref25}, (ii) high-scale video courses require interaction designs that sustain attention and prompt response \citep{ref21,ref22,ref23}, (iii) tutoring human or intelligent produces medium-to-large learning gains relative to business-as-usual instruction \citep{ref19,ref26}, and (iv) test-enhanced learning strengthens long-term retention and transfer, especially with spaced retrieval \citep{ref27}. By combining experimental rigor with scalable deployment, this pilot study provides preliminary evidence suggesting that conversational AI tutors may enhance video-based education while preserving accessibility and reach \citep{ref1,ref3,ref11,ref17,ref19,ref20}.

\section{Related Work}

The foundation of modern online education rests on video-based content, a format that has proven highly effective for content delivery at scale. Platforms like Coursera and edX have demonstrated the global reach of MOOCs, but also their inherent limitations. Research by \citet{ref21} showed that a vast majority of learners on MOOC platforms watch only a fraction of each video, with engagement dropping off sharply after the first few minutes. This finding highlights a fundamental challenge: passive video viewing often fails to maintain attention, leading to superficial learning and high dropout rates \citep{ref22,ref23}.

To address the limitations of passive learning, the field of Intelligent Tutoring Systems (ITS) has sought to create personalized, interactive learning environments. Early ITS models provided structured problem sets and rule-based feedback, demonstrating significant learning gains comparable to one-on-one human tutoring \citep{ref19}. More recently, the advent of Large Language Models (LLMs) has revolutionized this field by enabling more naturalistic, conversational interactions. Studies have found that LLM-powered conversational agents can answer complex questions, offer hints, and provide explanations in a way that mimics human-to-human dialogue \citep{ref8,ref18}. These systems are particularly adept at fostering deeper questioning, a key indicator of cognitive engagement \citep{ref13}.

However, a critical gap remains between these promising small-scale studies and their application in real-world, large-scale educational settings. While some platforms have experimented with chatbots, these are often supplemental or limited in their pedagogical scope \citep{ref9,ref11}. The core instructional format, the video, remains largely unchanged. Our work is one of the first to systematically investigate a seamless, pedagogically-aligned integration of conversational AI directly into the video lecture format itself. By combining the scalability of traditional media with the adaptive power of real-time dialogue, our platform aims to bridge the chasm between passive consumption and the active, constructive learning behaviors necessary for mastering complex topics like AI.

\section{Methods}

\subsection{Participants and Design}

Our study employed a within-subjects experimental design with a total of 58 participants (N = 58) recruited from a diverse global population, with a mean age of M = 21.4 years (SD = 2.8). Participants represented varied academic backgrounds and career stages: Working Professionals (37.9\%), 4th Year Undergraduates (22.4\%), Postgraduates (15.5\%), 3rd Year Undergraduates (15.5\%), PhD Scholars (5.2\%), and 2nd Year Undergraduates (3.4\%). The majority of participants had backgrounds in Computer Science, AI, or ML (81.0\%), with smaller representations from Other Engineering fields (6.9\%), Electrical/Electronics Engineering (3.4\%), and other disciplines (8.6\%). Participants were geographically distributed across 49 unique locations globally, with concentrations in major technology hubs. All participants were multilingual, with English proficiency being universal across the sample.

\subsection{Learning Conditions}

Participants were tasked with learning a fundamental concept in AI: Tokenization. The instructional content volume for each condition was standardized, utilizing a 60-minute video lecture. While the source material was identical in length, actual participation time was variable as students were permitted to navigate the content at their own pace.

\textbf{Condition A (Traditional Video):} The instructional content for this condition consisted of a 60-minute pre-recorded video lecture that explained the concept of tokenization using static visuals and spoken narration. While the total available content was standardized at 60 minutes, participants were permitted to progress through the material at their own pace, including the use of playback speed controls.

\textbf{Condition B (AI-Augmented Video):} Participants engaged with the same 60-minute video lecture content. However, the video interface was augmented with a real-time conversational AI tutor powered by a large language model architecture. This tutor was accessible via a chat window that appeared alongside the video player. The AI tutor implemented several key pedagogical features:

\textit{Proactive Intervention:} The system automatically identified key junctures in the video content (concept transitions, complex explanations, and example presentations) through pre-programmed timestamps and triggered contextual questions to prompt active recall and deeper reflection.

\textit{Adaptive Formative Assessment:} The tutor provided personalized quizzes and knowledge checks based on the current video segment, adjusting difficulty based on student responses and offering immediate explanatory feedback.

\textit{Conversational Scaffolding:} Students could engage in natural language dialogue to request clarification, explore related concepts, or receive step-by-step problem-solving guidance. The tutor maintained conversation context throughout the learning session.

\textit{Misconception Detection:} The system was designed to identify common student misconceptions about tokenization and provided targeted interventions with alternative explanations and corrective examples.

\subsubsection{Technical Implementation and Temporal Grounding}

The platform's conversational engine was powered by the Gemini 1.5 Flash (gemini-1.5-flash-002) model via the Multimodal Live API. This choice was motivated by the model's native support for low-latency, voice-to-voice interaction and its 1-million-token context window. To ensure precise temporal grounding and minimize hallucinations, we implemented a Timeline-Augmented Retrieval-Augmented Generation (RAG) architecture.

The lecture video was first processed into discrete segments. For each segment, a concise pedagogical summary and key metadata (including technical terms like ``BPE'' or ``subword units'') were generated and stored as vectorized embeddings using the text-embedding-004 model. During student interaction, the system dynamically retrieves the specific context corresponding to the current video timestamp. This allows the API to provide highly localized and accurate feedback based on the exact visual and verbal content the student is viewing at that moment. The model was configured with a temperature of 0.3 to prioritize factual consistency and a max output limit of 512 tokens to maintain a concise, Socratic tutoring style.

Misconception detection was implemented via Chain-of-Thought prompting rather than fine-tuning. The system was provided with a curated list of common NLP errors (e.g., confusing BPE with WordPiece tokenization, misunderstanding vocabulary size constraints). When a user query matched the semantic embedding of a known misconception, the LLM was prompted to provide a Socratic correction rather than a direct answer, encouraging the learner to reason through the error independently. This prompt-based approach allowed for flexible adaptation without requiring model retraining.

\begin{figure}[ht]
\centering
\includegraphics[width=\columnwidth]{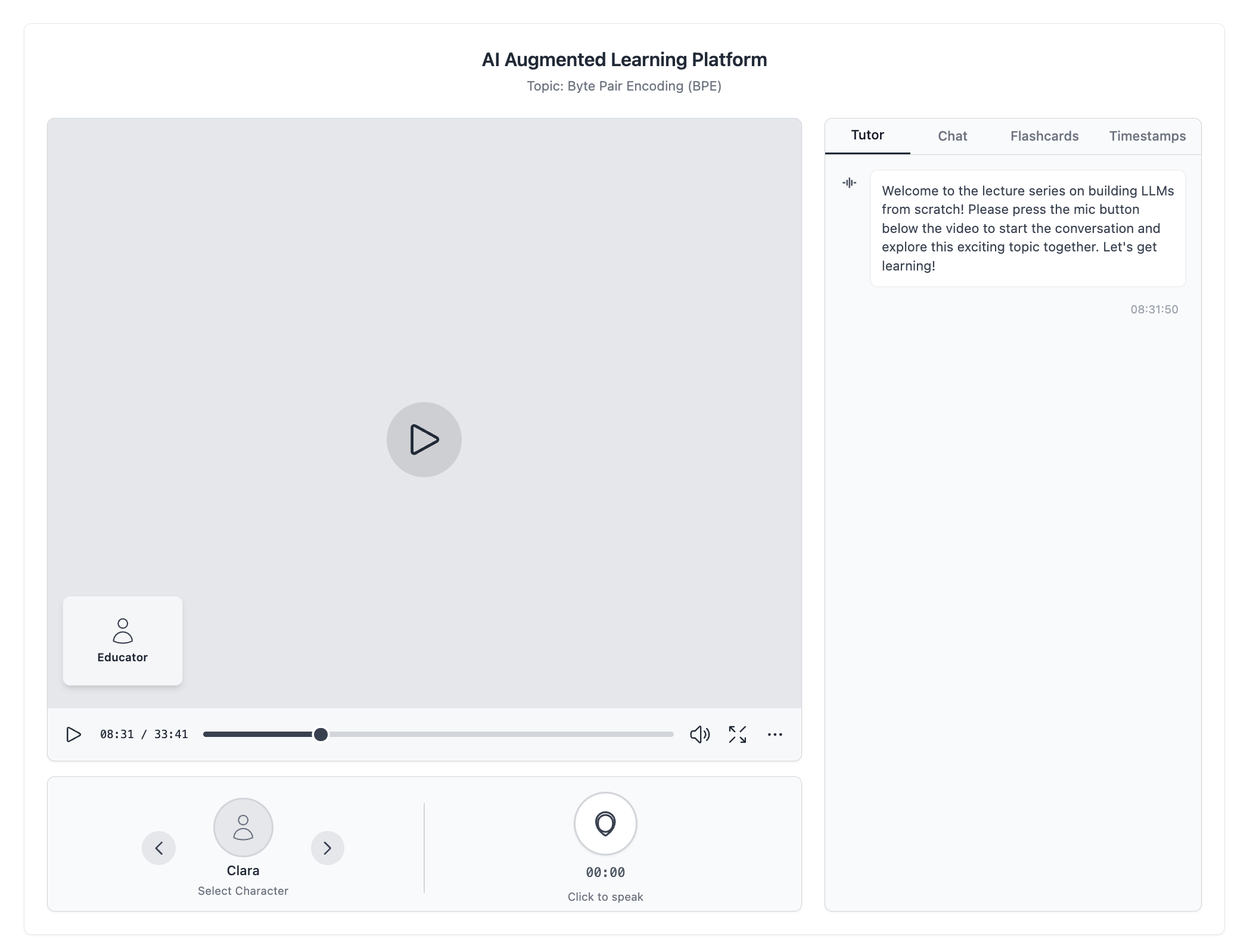}
\caption{The AI Augmented Learning Platform interface demonstrating the experimental setup for Condition B. The platform integrates a video player (showing Byte Pair Encoding topic) with an interactive tutor chat system, character selection features, and voice interaction capabilities. The tutor provides real-time guidance and welcomes learners to engage actively with the content.}
\label{fig:tutor_interface}
\end{figure}

\subsubsection{Study Procedure and Instructions}

\textbf{Condition A -- Traditional Video:} All participants first engaged with a pre-recorded video lecture on tokenization (60 minutes of total content) followed by an immediate assessment (Immediate Pre-test). Participants were allowed to control their viewing experience, including playback speed and navigation. This served as both a baseline measure and a learning opportunity that provided participants with initial exposure to the content.

\textbf{Condition B -- AI-Augmented Learning:} Participants then engaged with the same video content enhanced with an interactive AI tutor, followed by an Immediate Post-test. This design provided a conservative assessment of AI tutor effectiveness, as any observed improvements represent a lower-bound estimate given the potential contribution of repeated content exposure.

Participants were also given specific guidelines to ensure the integrity of the research. They were instructed to provide honest answers and to refrain from searching for information online. The study was not presented as a competition, and participants were assured that their scores would be used for research purposes only. To optimize the experience, a laptop or PC screen was required for participation.

\begin{figure*}[ht]
\centering
\includegraphics[width=0.8\textwidth]{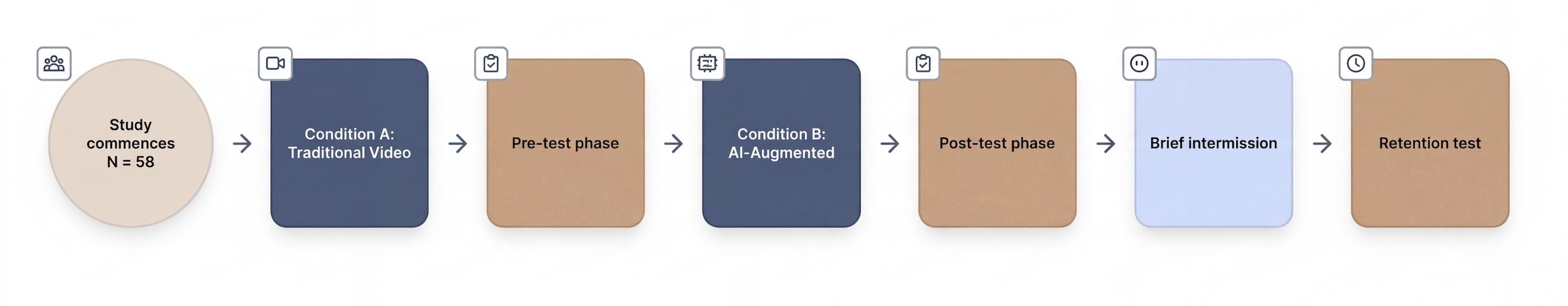}
\caption{Experimental procedure flow diagram showing the sequential within-subjects design used in this study. All participants (N=58) progressed through sequential conditions: Condition A (Traditional Video) with Immediate Pre-test, Condition B (AI-Augmented) with Immediate Post-test, brief intermission, and Delayed Retention Assessment. This design enabled direct within-subjects comparisons of learning outcomes.}
\label{fig:procedure_flow}
\end{figure*}

\subsection{Measures}

We measured learning outcomes and engagement patterns using three distinct assessments and behavioral analytics. We refer to the assessment following Condition A as an `Immediate Pre-test' relative to AI augmentation, and the assessment following Condition B as the `Immediate Post-test'.

\textbf{Immediate Post-test:} A 10-item multiple-choice test and 3 open-ended questions administered immediately after each condition to assess basic knowledge and conceptual recall. Multiple-choice questions were scored for accuracy, while the open-ended questions were graded on a rubric for a total possible score.

\textbf{Delayed Retention Assessment (2-week delay):} A re-administration of the same 10-item multiple-choice test and 3 open-ended questions, two weeks after the completion of Condition B (AI-Augmented Learning), to measure long-term knowledge retention. The same scoring procedures were applied.

\textbf{Behavioral Analytics:} We logged comprehensive user interactions including session duration, chat frequency, question types, tutor engagement patterns, and time-on-task metrics. Privacy considerations were addressed through data anonymization and informed consent procedures, with all interaction logs stripped of personally identifiable information. The complete anonymized dataset containing participant responses, test scores, and engagement metrics is provided as supplementary material to support reproducibility.

\subsection{Data Analysis}

We used paired-samples t-tests to compare learning outcomes between the pre-test (traditional video) and post-test (AI-augmented) conditions. Cohen's d effect sizes were calculated for each comparison using the formula:

\begin{equation}
d = \frac{M_1 - M_2}{SD_{pooled}}
\end{equation}

where $SD_{pooled} = \sqrt{\frac{(n_1-1)SD_1^2 + (n_2-1)SD_2^2}{n_1+n_2-2}}$

This standardized effect size measure quantifies the magnitude of difference between conditions, with values of 0.2, 0.5, and 0.8 typically interpreted as small, medium, and large effects, respectively.

\textbf{Design Rationale:} We chose a within-subjects sequential design over a between-subjects A/B test for three key reasons. First, given the high diversity in our sample (N=58, ranging from undergraduate students to PhD scholars and working professionals with varying NLP backgrounds), a within-subjects design ensures that each participant serves as their own control, eliminating noise caused by baseline differences in NLP knowledge that a between-subjects design with only 29 participants per group might not adequately balance. Second, within-subjects designs offer increased statistical power compared to between-subjects designs of equivalent sample size by removing individual variation from the error term, enabling more sensitive detection of the AI tutor's pedagogical effects. Third, as an exploratory pilot study, a fixed sequential order was utilized rather than counterbalancing to maintain the pedagogical scaffolding of the curriculum. Since the AI-augmented condition (Condition B) was designed to build upon the baseline conceptual framework established in the traditional video, presenting the AI condition first would have introduced significant confounding variables regarding a participant's ability to engage with the tutor without prior foundational exposure. This sequential approach allowed us to observe complete learning trajectories while ensuring the instructional logic remained coherent for the learner.

We also computed Pearson correlations to examine relationships between engagement metrics and learning gains. Linear regression models were fitted to predict post-test performance from pre-test scores and to examine the relationship between engagement changes and learning improvements. All analyses acknowledge the limitation that observed differences may reflect both AI tutor effects and practice/familiarity effects due to the fixed-order design.

\section{Results}

Our analysis revealed significant improvements in learning outcomes and engagement behaviors for participants using the AI-augmented platform compared to traditional video-based instruction.

\subsection{Learning Outcomes}

As hypothesized, the AI-augmented condition resulted in significantly higher scores across all three assessments. Figure 3 illustrates the difference in mean scores across assessment types, while Figure 4 shows individual learning trajectories. Table 1 provides the detailed statistical results.

\begin{figure}[ht]
\centering
\includegraphics[width=\columnwidth]{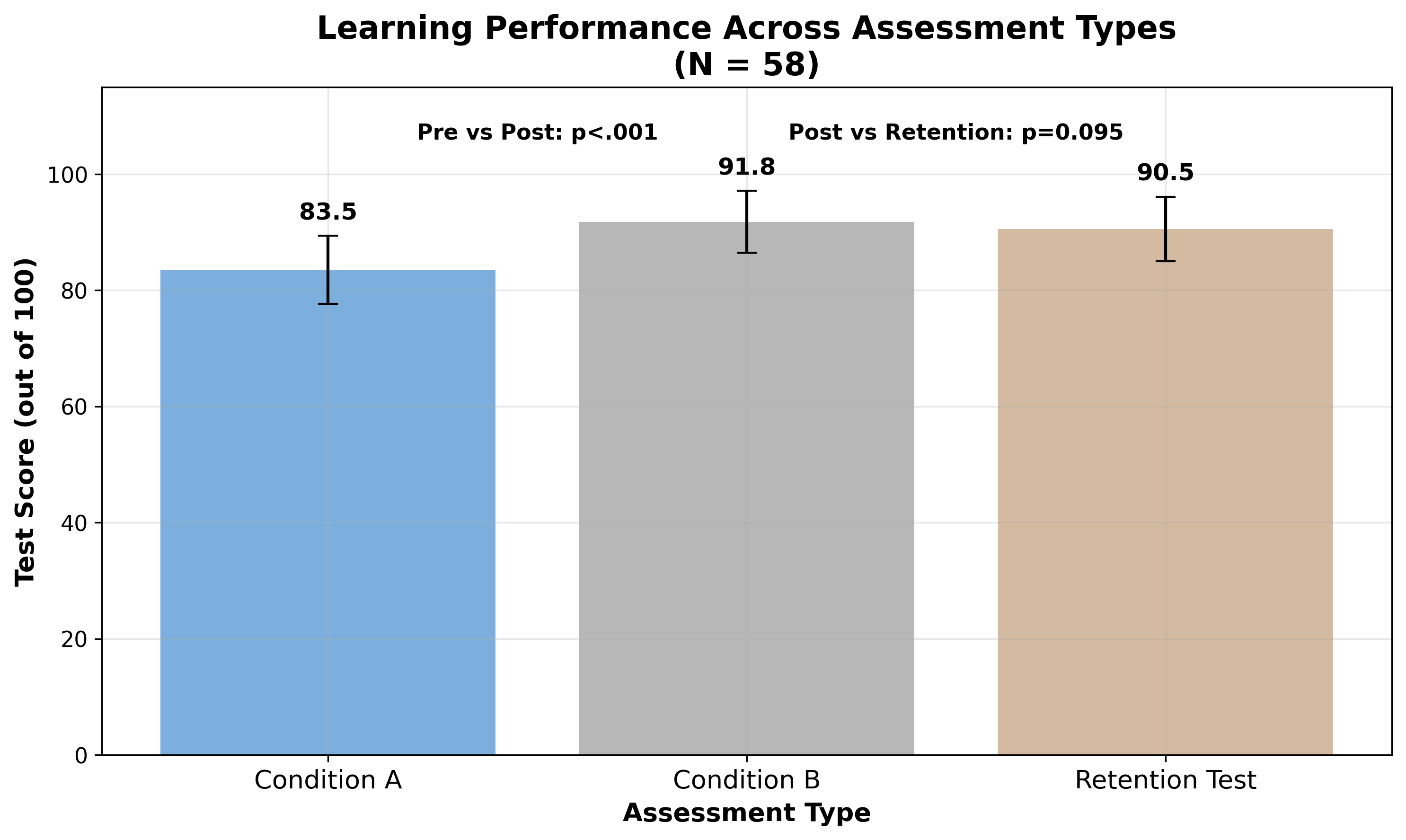}
\caption{Learning performance comparison across assessment types showing significant improvement from pre-test to post-test ($p < .001$) with maintained performance in retention testing. Scores are out of 100 points. Error bars represent standard deviation.}
\label{fig:learning_performance}
\end{figure}

\begin{figure}[ht]
\centering
\includegraphics[width=\columnwidth]{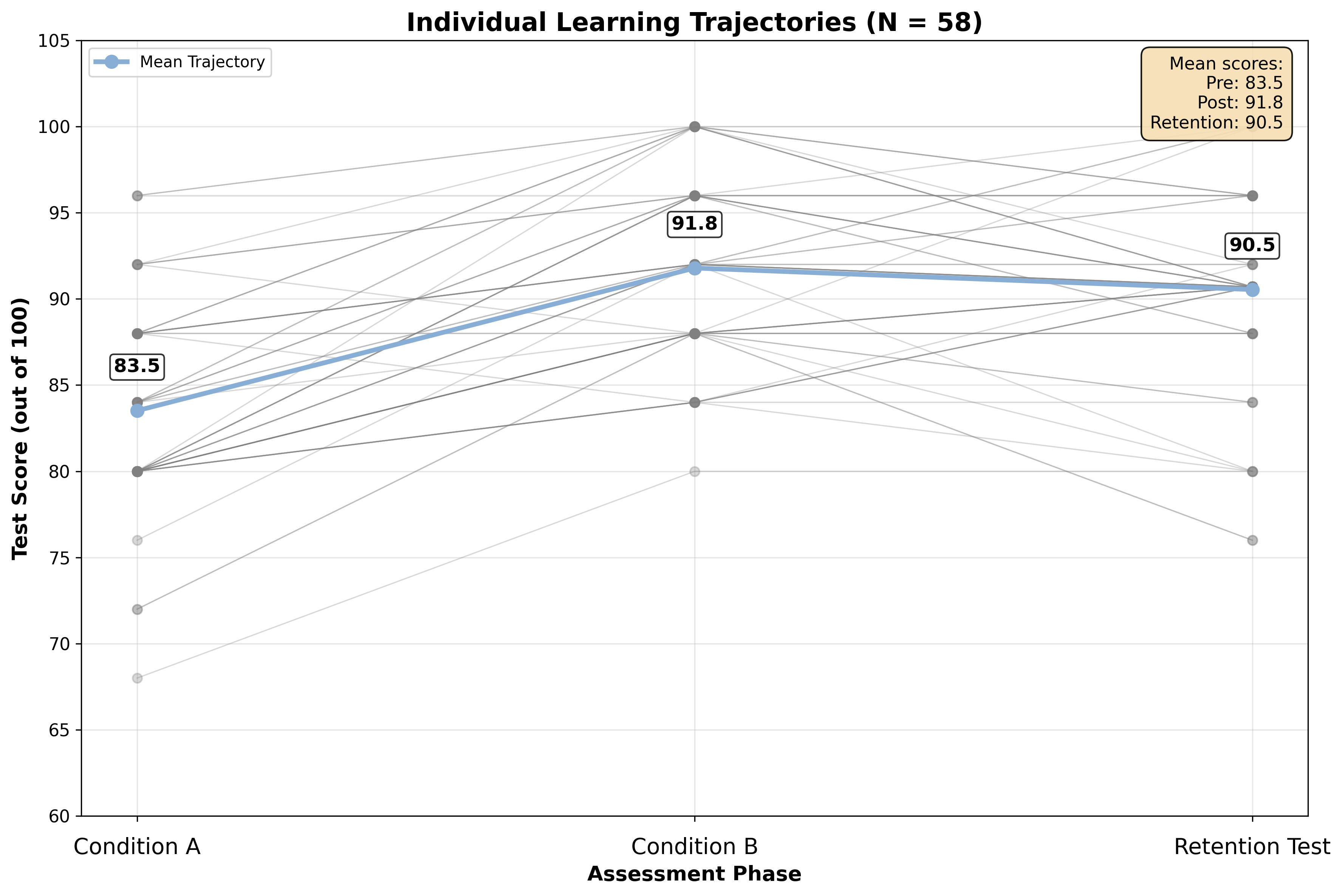}
\caption{Individual learning trajectories across all participants (N=58) showing varied patterns of improvement. The red line represents the mean trajectory, demonstrating overall learning gains from pre-test to post-test with slight decline in retention testing. Scores are out of 100 points.}
\label{fig:learning_trajectories}
\end{figure}

\begin{table*}[ht]
\centering
\small
\begin{tabular}{lccccc}
\hline
\textbf{Assessment} & \textbf{Condition B} & \textbf{Condition A} & \textbf{t(57)} & \textbf{p-value} & \textbf{Cohen's d} \\
 & \textbf{(AI-Augmented)} & \textbf{(Traditional Video)} & & & \\
\hline
Immediate Post-test & 91.8 $\pm$ 5.3 & 83.5 $\pm$ 5.9 & -11.462 & $p < .001$ & 1.505 \\
\\[0.5em]
Delayed Retention Assessment & 90.5 $\pm$ 5.6 & N/A & N/A & N/A & N/A \\
\hline
\end{tabular}
\caption{Comparison of Learning Outcomes. Condition B = AI-Augmented, Condition A = Traditional Video. Scores out of 100. Paired-samples t-tests (N=58).}
\label{tab:learning_outcomes}
\end{table*}

The immediate post-test showed a statistically significant difference, with the AI-augmented group scoring 8.3 points higher on average (91.8 vs 83.5 out of 100; t(57)=-11.462, $p < .001$). The effect size of d=1.505 indicates a large impact. The delayed retention assessment, administered two weeks later, showed maintained performance with participants retaining knowledge effectively (90.5 out of 100), with a slight non-significant decline from immediate post-test (t(57)=1.695, $p = .095$). The learning trajectory analysis revealed consistent improvements across the majority of participants, demonstrating the effectiveness of AI-augmented learning.

\subsection{Behavioral and Engagement Analytics}

Behavioral data revealed substantial differences in engagement patterns between conditions. The observed average duration of 31.8 minutes in Condition A (SD = 10.7) highlights the `passive-skipping' behavior typical of traditional video learning: while 60 minutes of content was available, platform logs confirmed students optimized for speed using playback controls (1.5$\times$ or 2$\times$ speed) and skipped familiar segments. In contrast, the AI tutor in Condition B successfully re-engaged learners for a significantly longer duration (M = 54.5 minutes, SD = 19.2), representing a 71.1\% increase in time-on-task. Interaction logs revealed approximately 40\% of this session time was spent actively engaging with the AI tutor (asking questions, responding to prompts), while 60\% involved passive viewing, demonstrating a fundamental shift from passive consumption to active dialogue-based learning. Figure 5 illustrates the engagement duration comparison. Table 2 summarizes behavioral findings.

\begin{figure*}[ht]
\centering
\includegraphics[width=0.9\textwidth]{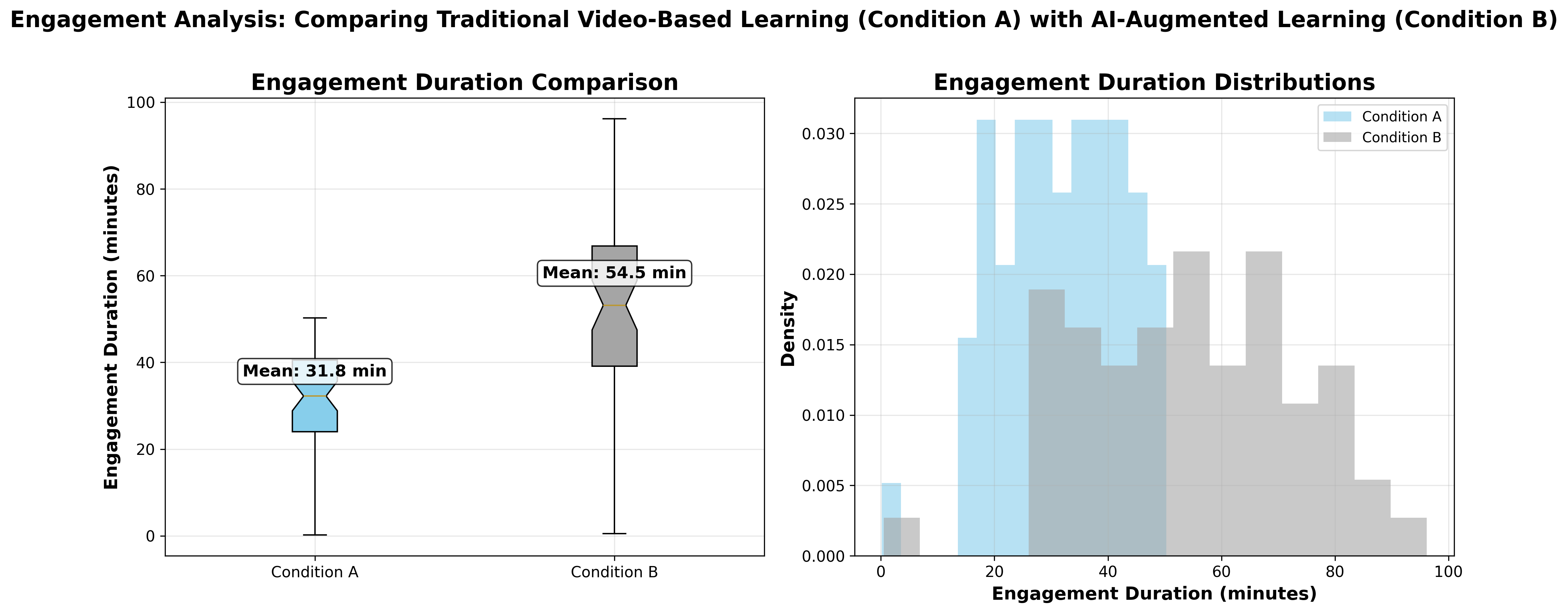}
\caption{Engagement duration analysis comparing traditional video-based learning (pre-test) with AI-augmented learning (post-test). Left panel shows box plot comparison with mean values. Right panel shows density distribution revealing significantly increased engagement with AI tutoring.}
\label{fig:engagement_analysis}
\end{figure*}

\begin{table}[ht]
\centering
\small
\begin{tabular}{lcc}
\hline
\textbf{Metric} & \textbf{Condition A} & \textbf{Condition B} \\
 & \textbf{(Traditional Video)} & \textbf{(AI-Augmented)} \\
\hline
Avg Session & 31.8 $\pm$ 10.7 min & 54.5 $\pm$ 19.2 min \\
Duration & & \\
\\[0.3em]
Chat & N/A & 23.4 $\pm$ 6.8 msgs \\
Interactions & & \\
\\[0.3em]
Proactive & N/A & 8.7 $\pm$ 2.1 resp \\
Questions & & \\
\\[0.3em]
Session Duration & Baseline & +71.1\% \\
Increase & & \\
\hline
\end{tabular}
\caption{Behavioral Analytics of Learner Engagement. Condition A = Traditional Video, Condition B = AI-Augmented. Values show mean $\pm$ SD.}
\label{tab:behavioral_analytics}
\end{table}

\section{Discussion}

The results of this pilot study provide preliminary evidence suggesting that integrating conversational AI tutors into traditional video-based instruction may enhance the learning process. The gains observed across immediate post-test performance and long-term retention support the hypothesis that active, dialogic engagement outperforms passive content consumption. The large effect sizes align with prior research on the efficacy of intelligent tutoring systems, and our pilot study demonstrates this potential effect within a highly prevalent learning format.

The behavioral analytics offer a crucial insight into the mechanisms behind these gains. The 71.1\% increase in average session time suggests that the AI tutor served as a dynamic cognitive partner, encouraging learners to process information more deeply. Rather than just watching a video, participants were actively constructing knowledge through questioning, problem-solving, and immediate feedback. This aligns with constructivist theories of learning and the well-established benefits of test-enhanced learning, where the act of retrieving and applying knowledge strengthens memory and transfer \citep{ref27}.

\textbf{Design Considerations:} Our sequential within-subjects approach, while providing direct comparisons, means the observed 8.3-point improvement may reflect both AI tutor effects and practice effects. The substantial effect size (d=1.505) and 71\% increase in engagement duration suggest meaningful benefits that warrant further investigation beyond content repetition alone. Together, these behavioral and performance shifts indicate practical promise for AI-augmented learning, while motivating more rigorous controlled evaluations.

\textbf{Scope and Generalizability:} The study demonstrates effectiveness for tokenization instruction within AI education, establishing a foundation for broader investigation across diverse educational domains. Our controlled setting enabled precise measurement of learning outcomes while maintaining experimental rigor.

\textbf{Sample and Duration:} Our university student sample (N=58, ages 19-25) provided valuable insights into AI-augmented learning for this demographic, with the 2-week retention period confirming sustained learning benefits.

\subsection{Future Research Directions}

\textbf{Enhanced Experimental Designs:} Future studies could employ randomized controlled trials with between-subjects designs or counterbalanced condition orders to further validate and quantify AI tutor effectiveness across diverse learning contexts.

Additional research directions include longitudinal studies examining knowledge retention over months rather than weeks to understand the durability of AI-augmented learning gains. Cross-domain validation studies should test the platform's effectiveness across STEM and non-STEM subjects to establish broader applicability.

Qualitative analysis of student-AI conversations could illuminate optimal dialogue patterns, identify common student misconceptions, and inform improvements to the AI tutor's pedagogical strategies. Additionally, research should explore individual difference factors such as learning styles, prior technology experience, and metacognitive awareness to develop more personalized AI tutoring approaches.

\textbf{Scalability to Complex NLP Concepts:} The Timeline-Augmented RAG architecture is inherently designed to generalize across various NLP domains. For procedural or technical topics---such as parsing algorithms, neural architectures, or multilingual preprocessing---the system provides high-fidelity grounding by mapping student queries to specific instructional segments. However, scaling to more abstract concepts like pragmatics or discourse analysis presents a unique challenge: these topics often require a `global' understanding of context that spans the entire lecture rather than a single timestamped `local' segment. Future iterations of the platform will explore a hierarchical RAG approach, combining local timestamped metadata with a global thematic discourse map. This would allow the AI to facilitate deeper theoretical debates (e.g., compositional vs. distributional semantics) without the risk of oversimplification, ensuring the tutor remains a sophisticated pedagogical partner as curriculum complexity increases.

\section{Conclusion}

This pilot study provides preliminary evidence that the integration of conversational AI tutors into video-based education may improve learning outcomes and engagement. By transforming a passive viewing experience into an interactive and adaptive dialogue, our hybrid learning platform showed enhanced knowledge retention and improved ability to transfer knowledge to new contexts. The findings suggest a potential avenue for scalable, effective online education that leverages the cost-effectiveness of video while addressing its pedagogical shortcomings. The conversational AI tutor, powered by the latest advances in large language models, represents a promising tool for educators, with potential to provide personalized, on-demand support to learners at scale. Future research with randomized controlled designs is needed to confirm these preliminary findings and establish causal relationships.

\section*{Limitations}

While our study demonstrates significant improvements in learning outcomes with AI-augmented video instruction, several limitations should be acknowledged. First, the sequential within-subjects design means that all participants experienced the traditional video condition before the AI-augmented condition, which may have contributed to the observed improvements through practice effects. While large effect sizes and engagement shifts were observed, the fixed-order design prevents isolating the causal contribution of the AI tutor. These results should therefore be interpreted as preliminary signals motivating controlled follow-up studies.

Second, our study focused on a single educational domain (tokenization in AI) with a specific demographic (university students aged 19-25). While this provided precise measurement of learning outcomes within a controlled setting, broader validation across diverse educational domains and age groups is needed to establish generalizability.

Third, the 2-week retention period, while demonstrating sustained learning benefits, represents a relatively short timeframe for assessing long-term knowledge retention. Future work should examine knowledge persistence over months or years.

Finally, our quantitative metrics captured substantial improvements in engagement duration and learning performance, but qualitative analysis of student-AI conversational transcripts could provide deeper insights into optimal tutoring strategies and common misconceptions that warrant further investigation.

\section*{Ethical Considerations}

The implementation of AI tutoring systems raises important ethical considerations that warrant careful attention. Student data privacy emerged as a primary concern, particularly given the collection of detailed interaction logs, conversation transcripts, and learning analytics. Our study addressed these concerns through comprehensive data anonymization procedures and informed consent protocols that clearly outlined data usage and retention policies.

However, broader deployment of such systems must consider potential algorithmic bias in AI responses, ensuring that the tutoring system provides equitable support across diverse student populations. Future implementations should include bias detection mechanisms and regular audits of AI tutor responses for fairness and accuracy.

Additionally, concerns about over-reliance on AI tutoring systems must be addressed. While our study showed positive learning outcomes, educators should maintain awareness of the importance of human interaction and critical thinking skills that may not be fully supported by AI systems alone.

\bibliographystyle{unsrtnat}
\bibliography{references}

\end{document}